\documentclass{moriond}

\def\be{\begin{equation}}
\def\ee{\end{equation}}
\def\bea{\begin{eqnarray}}
\def\eea{\end{eqnarray}}

\newcommand{\Photo}{}

\usepackage{amsmath}
\usepackage{amssymb}

\begin{document}
\vspace*{4cm}

\title{NEUTRINO DECOUPLING AND THE TRANSITION TO COLD DARK MATTER}

\author{ ROMAN SCHNABEL }

\address{Institut f\"ur Laserphysik \& Zentrum f\"ur Optische Quantentechnologien, Universit\"at Hamburg, \\
Luruper Chaussee 149, 22761 Hamburg, Germany}

\maketitle
\abstracts{
About 40 years ago, the neutrino was ruled out as the dark matter particle based on several arguments.
Here I use the well-established concept of quantum uncertainties of position and momentum to describe the decoupling of neutrinos from the primordial plasma, which took place about half a second after the Big Bang. In this way I show that the main arguments against the neutrino are either wrong or have loopholes, and conclude that the neutrino urgently needs to be reconsidered, not as a `hot', but as the `cold' dark matter particle.
}

\section{Introduction} \vspace{-2mm}
One of the greatest mysteries in astrophysics and cosmology is the nature and the origin of cold dark matter, which makes up more than 84\% of the mass in the Universe. 
Dark matter creates and reacts to gravity and determines the dynamics of stars around galactic centres, however it does not absorb or emit electromagnetic radiation. 
As early as the 1930s, the existence of dark matter was inferred from the movements of galaxies in galaxy clusters, and it was posited that most of the matter in the Universe could be dark \cite{DeSwart2017}.\\ 
A candidate for the particle of cold dark matter should have non-zero rest mass and be neutral but not baryonic.
In the late 1970s, the neutrino seemed to be a likely candidate, but already in the early 1980s, the neutrino was essentially excluded as the dark matter particle \cite{Tremaine1979,White1983,Kolb1990}. 
First, neutrinos are fermions, and the Pauli exclusion principle sets an upper bound to the number density for neutrinos with a given momentum spread. 
This consideration results in a lower bound for the particle mass, which is orders of magnitude higher than the estimated neutrino mass \cite{Tremaine1979,Kolb1990}. 
Second, the neutrino as the dark matter particle lost more power of persuasion when computer simulations showed that the highly abundant, ultra-relativistic neutrinos are too `hot' to produce the observed scale of galaxy clustering \cite{White1983}. The actual dark matter particle needs to be `cold'. 
The third argument against the neutrino as the dark matter particle is no less vital. Due to the expansion of the Universe, the neutrinos should have basically lost their relativistic mass. The rest mass of all neutrinos in the Universe, however, is about two orders of magnitude lower than the mass of the dark matter. 
The neutrino has thus been ruled out. Unfortunately, there are no other known particles that suit as the dark matter particle.
 The search for novel, hypothetical particles began many years ago, but has so far not been successful \cite{Sikivie1983,Goodman1985,Ahmed2011,Aprile2012,Akerib2014,Bertone2018}.\\ \vspace{-8mm}

\section{The neutrinos at decoupling} \vspace{-2mm}
Neutrino decoupling took place about half a second after the Big Bang ($t_{\nu {\rm d}} = 0.5$\,s), when the thermodynamic temperature of the Universe was about $T_{\nu {\rm d}} \approx 3 \cdot 10^{10}$\,K\, \cite{deSalas2016}. 
The term `decoupling' describes a rather instantaneous process, after which the collision rate between a neutrino and any other particle was basically zero. Neutrino decoupling was caused by the expansion of the Universe and the associated reduction of energy density and thermodynamic temperature \cite{Sciama1994}.
At decoupling temperature $T_{\nu {\rm d}}$, the (weak-force) interaction rate of the neutrinos 
dropped below the relative one-dimensional expansion rate of the Universe, 
which meant that the interval between two momentum-changing interactions started to exceed the age of the Universe \cite{Sciama1994}. 

Whether the neutrinos at decoupling had to be considered as a Fermi gas rather than a Maxwell-Boltzmann gas can be estimated by comparing the average neutrino distance $d$ as derived by the (fully classical) Boltzmann equation with twice the standard deviation of the Gaussian position uncertainty $2\Delta \hat x$. The average one-dimensional distance between two identical neutrinos reads \cite{Sciama1994,Pastor2011}
\vspace{-3mm}
\begin{equation} \label{eq:d}
\hspace{8mm}
d \approx \sqrt[3]{\frac{1}{6 \pi \cdot \zeta(3)}} \cdot \frac{h c}{k_B T_{\nu {\rm d}}} \approx 0.35 \cdot \frac{h c}{k_B T_{\nu {\rm d}}} \approx 1.7 \cdot 10^{-13}\,{\rm m} \;,
\end{equation}
with $\zeta(3) \approx 1.2$ and $h$ is the Planck constant, $c$ is the speed of light, and $k_B$ is the Boltzmann constant. The thermal de\,Broglie wavelength for ultra-relativistic particles, for which collision rates can be neglected compared to the annihilation and creation rates, reads $\lambda_{\rm th}  \approx hc/(k_B T)$ for one dimension. Half this wavelength corresponds to the smallest particle position spread. Dividing by $\pi$ approximates the standard deviation of a Gaussian uncertainty, i.e.~$\Delta \hat x \approx \lambda_{\rm th} / (2 \pi)$, which yields
\vspace{-2mm}
\begin{equation}
\label{eq:2dx}
2 \Delta \hat x_{\nu{\rm d}} \approx \frac{\lambda_{{\rm th} \nu {\rm d}}}{\pi} = \frac{1}{\pi} \cdot \frac{h c}{k_B T_{\nu {\rm d}}} \approx 0.32 \cdot \frac{h c}{k_B T_{\nu {\rm d}}} \approx 1.5 \cdot 10^{-13}\,{\rm m} \;.
\vspace{1mm}
\end{equation}
The almost identical values imply that the neutrinos at decoupling were at the transition to a quantum degenerate Fermi gas. 
The neutrino \emph{momentum} distribution followed the Fermi-Dirac statistic for relativistic particles. According to this, the average momentum is a good approximation for the distribution's full width$^{\rm(fw)}$ $\langle p \rangle_{\rm th}  \approx  \Delta^{\rm fw}_{\rm th} p \approx k_B T / c$, see for instance Fig.\,1 in Ref.\,\cite{Liu2012}. It turns out that the neutrinos at decoupling do \emph{not} have different classes of momenta because the entire momentum spread is given by quantum uncertainty, and the position and momentum uncertainties of the neutrinos at decoupling correspond to a \emph{pure} kinetic state, since the uncertainty product is close to the lower bound of the Heisenberg uncertainty relation \cite{Robertson1929}
\vspace{0mm}
\begin{equation}
\label{eq:up}
\Delta \hat x_{\nu{\rm d}} \cdot \Delta \hat p_{\nu{\rm d}} \approx  \frac{\lambda_{{\rm th,} \nu {\rm d}}}{2\pi} \cdot  \frac{\Delta^{\rm fw}_{\rm th} p_{\nu{\rm d}}}{2} \approx  \frac{1}{2\pi} \frac{h c}{k_B T_{\nu {\rm d}}} \cdot \frac{k_B T_{\nu {\rm d}}}{2 c} = \frac{h}{4 \pi}\;.
\end{equation}
The meaning of the finding in Eq.\,(\ref{eq:up}) is the following. When the neutrinos decoupled, almost their entire momentum spread was given by quantum uncertainty. The momentum of every single neutrino was not directed and had expectation value zero. Due to the spherical momentum uncertainty, the position uncertainty increased in a spherical way. It can be shown that in turn the momentum uncertainty reduces during this expansion, keeping the product of the position and momentum uncertainties at the minimum of the Heisenberg uncertainty relation. (A simple argument is that the product is a monotonically increasing function of the kinetic temperature, which cannot increase due to energy conservation.)

\section{The neutrino interference after decoupling} \vspace{-1mm}
After decoupling, the free evolution of the Gaussian neutrino modes was fundamentally different from the classical concept of free, particle-like propagation. The free evolution of a neutrino mode was described by the Schr{\"o}dinger equation with a flat potential (of quasi-infinite expansion): the position uncertainties increased almost at the speed of light due to the ultra-relativistic momentum uncertainties. 
The inevitable consequence was mode overlap and the emergence of neutrino indistinguishability, realised simply by interference in such a way that the Pauli exclusion principle was not violated. This resulted in a delocalised three-dimensional standing wave, a delocalised `neutrino crystal', in which nodes of the standing wave spatially separated the fermionic neutrinos. The wavelength of this `crystal', or Fermi sea, was just slightly longer than the thermal De\,Broglie wavelength at decoupling ($\lambda_{\rm Fs} \gtrsim \lambda_{{\rm th},\nu d}$). In the nodes, the probability of finding a neutrino was zero. The locations of nodes and neutrinos, however, were not determined with respect to the primeval plasma. Most importantly, all neutrinos were smeared out over the entire dimension of the emerged macroscopic quantum system and indistinguishable, but their uncertainty had internal quantum correlations in such a way that residual decoherence, e.g.~due to a collision with an electron of the primeval plasma, never localised two identical neutrinos in the same antinode.

\section{Gravity-supported condensation and Cooper pairing}
\vspace{-1mm}
At decoupling, the neutrinos' kinetic uncertainties corresponded to an almost pure state. Their momentum expectation value was extremely close to zero, and their kinetic energy was exclusively due to momentum uncertainty. While the position uncertainties subsequently increased, the momentum uncertainty naturally decreased, maintaining the pure state characteristics. While the neutrino crystals reached galactic dimensions, the entropy of the neutrinos as well as their kinetic temperature naturally approached zero. The mass of the neutrinos, however, had to maintain relativistic and was given by
\vspace{-1mm}
\begin{equation}
\label{eq:m}
m_{\rm rel} = h/(c \lambda_{\rm Fs}) \lesssim h/(c \lambda_{{\rm th},\nu d}) \ggg m_{1,2,3} \;,
\end{equation}
with $m_{1,2,3}$ being the three neutrino rest masses. 
The massive neutrino field obviously had to sense its own gravitational potential. 
Since the neutrino collision rate was already extremely close to zero, they condensed into the ground state of the Fermi sea.
I provide a quantitative description of the neutrinos' kinetic temperature after decoupling in Ref.\,\cite{Schnabel2020dm}.\\
Using the concept of pairing provides a complementary view on a neutrino Fermi sea, which is that of a Bose-Einstein condensate. Pairing of particles in a Fermi sea naturally occurs if there is a weak force of attraction between them. This is the assertion of the Bardeen, Cooper and Schrieffer (BCS) theory \cite{Bardeen1957}, which was formulated to describe the formation of Cooper pairs from initially distinguishable electrons \cite{Cooper1956} and superconductivity in metals at low temperatures. My proposal is that pairs of neutrinos can also be interpreted as bosons because of the gravitational potential of the macroscopic Fermi sea.

\section{The loophole in the `too low total neutrino mass' argument} \vspace{-1mm}
Today's average neutrino/anti-neutrino number density amounts to $3.4 \cdot 10^{8}$\,m$^{-3}$, taking into account two spin and three mass values \cite{Sciama1994,Pastor2011}. Using the upper bound for the average neutrino rest mass of $0.04\,{\rm eV}/c^2$ ($\approx 7 \cdot 10^{-38}$\,kg) \cite{Palanque-Delabrouille2015} the average rest-mass density is less than $2.4 \cdot 10^{-29}$\,kg/m$^3$. Any relativistic component of the neutrino mass seems negligible today due to the past expansion of the Universe. The average dark matter mass density is about a hundred times higher, namely about $2.3 \cdot 10^{-27}$\,kg/m$^3\,$ \cite{Komatsu2011}. This discrepancy has been a strong argument against the neutrino as the dominating dark matter particle.

The estimation of today's neutrino mass density, however, has a loophole for the following reason. It is very plausible, that the degenerate gigantic neutrino Fermi seas around the emerged super-massive black holes -- see Ref.\,\cite{Schnabel2020dm} -- acted locally against the expansion of the Universe. Only in locations free from dark matter, i.e.~in the gaps between the dark matter scaffolding, could the Universe `freely' expand. This might already be confirmed by observations since galaxy clusters are known to be stable against the expansion of the Universe \cite{Navarro1995}. My hypothesis would also explain the origin of the observed cosmic voids \cite{Geller1989}. To explain all cold dark matter by neutrino mass, their relativistic mass today needs to be about a hundred times higher than their average rest mass. In other words, the effective expansion of the `neutrino-field Universe' must have reduced the average relativistic energy in three dimensions from initially $3 \times 2.6$\,MeV to now about $3 \times 1.3$\,eV ($\approx 100 \times 0.04$\,eV), which corresponds to an average neutrino field red-shift of $z^\nu_{\nu {\rm d}} = 2.6\,{\rm MeV} / 1.3\,{\rm eV} \approx 2 \cdot 10^{6}$ instead of the literature value of $\tilde{z}^\nu_{\nu {\rm d}} =  T_{\nu,{\rm d}} / T_{\rm C \nu B}  \approx 1.5 \cdot 10^{10}$, where $T_{\rm C \nu B} = 1.945$\,K is the (then incorrect) value of the frozen temperature of the cosmic neutrino back ground \cite{Sciama1994}.

\section{Summary and conclusion}
\vspace{-1mm}
The prevailing description of neutrino decoupling half a second after the Big Bang gives a physical picture of neutrinos having trajectories. I argue that such a particle-like picture is incorrect.
After decoupling, according to my reasoning, the Gaussian neutrino wave-functions expanded spherically at almost the speed of light according to the Schr{\"o}dinger equation, overlapped, and evolved into gigantic degenerate neutrino fields in the ground state of their own gravitational potential. 
Following the BSC theory, I argue that the neutrinos in such a Fermi sea paired to bosons of zero spin, similar to Cooper pairing of electrons in superconducting metals. The nonclassical feature of the Fermi seas prohibited any kind of density fluctuation, which should have led to the emergence of primordial supermassive black holes without gravitational collapse \cite{Schnabel2020dm}. 
I argue that dark-matter fields around supermassive black holes experienced a much lower average expansion of space-time than electromagnetic radiation, since their gravitational attraction locally resisted against the expansion. If this is correct, today's neutrinos have a relativistic mass high enough to account for all of the cold dark matter in the Universe.

\section*{Acknowledgments}
\vspace{-2mm}
This work was supported by the European Research Council (ERC) project `MassQ' (Grant No.~339897). The author acknowledges Wilfried Buchm\"uller, Ludwig Mathey, Henning Moritz for helpful discussions and further acknowledges useful discussion within `Quantum Universe' (Grant No.~390833306), which is financed by the Deutsche Forschungsgemeinschaft (DFG, German Research Foundation) under Germany's Excellence Strategy - EXC 2121.


\vspace{0mm}
\section*{References}
\vspace{-2mm}
\begin{thebibliography}{99}

\bibitem{DeSwart2017} J.\,G. de Swart, G. Bertone, and J. van Dongen, 
\emph{Nat.\,Astron.} {\bf 1}, 59 (2017).

\bibitem{Tremaine1979} S. Tremaine and J.\,E. Gunn,  
\emph{Phys.\;Rev.\;Lett.} {\bf 42}, 407 (1979).

\bibitem{White1983} S.\,D.\,M. White, C.\,S. Frenk, and M. Davis, 
\emph{Astrophys.\;J.} {\bf 274}, L1 (1983).

\bibitem{Kolb1990} E.\,W. Kolb, \emph{The Early Universe}, p.\,373, CRC Press (1990).

\bibitem{Sikivie1983} P. Sikivie, 
\emph{Phys.\;Rev.\;Lett.} {\bf 51}, 1415 (1983).

\bibitem{Goodman1985} M.\,W. Goodman and E. Witten, 
\emph{Phys.\;Rev.\;D}  {\bf 31}, 3059 (1985).

\bibitem{Ahmed2011} Z. Ahmed \emph{et al.},  
\emph{Phys.\;Rev.\;D}  {\bf 84}, 011102 (2011).

\bibitem{Aprile2012} E. Aprile \emph{et al.}, 
\emph{Phys.\;Rev.\;Lett.} {\bf 109}, 181301 (2012.)

\bibitem{Akerib2014} D.\,S. Akerib, \emph{et al.}, 
\emph{Phys.\;Rev.\;Lett.} {\bf 112}, 091303 (2014).

\bibitem{Bertone2018} G. Bertone and T.\,M.\,P. Tait, 
\emph{Nature} {\bf  562},  51 (2018). 

\bibitem{Sin1994} S.-J. Sin, 
\emph{Phys.\;Rev.\;D}  {\bf 50}, 3650 (1994).

\bibitem{deSalas2016} P.\,F. de~Salas and S.~Pastor, 
\emph{J.\;Cosmol.\;Astropart.\;Phys.} {\bf 2016}, 051 (2016).

\bibitem{Sciama1994} D.\,W. Sciama, \emph{Modern Cosmology and the Dark Matter Problem}, CU Press (1994).

\bibitem{Pastor2011} S.~Pastor, 
\emph{ Phys.\;Part.\;Nucl.} {\bf 42}, 628 (2011).

\bibitem{Liu2012} F.-H. Liu, C.-X. Tian, M.-Y. Duan, and B.-C. Li, 
\emph{Adv.\;High\;Energy\;Phys.} {\bf 2012}, 1 (2012).

\bibitem{Robertson1929} H. P. Robertson, 
\emph{Phys.\;Rev.} {\bf  34}, 163 (1929).

\bibitem{Schnabel2020dm} R. Schnabel, 
arXiv:2006.11394 (2020).

\bibitem{Bardeen1957} J.~Bardeen, L.~N. Cooper, and J.\,R. Schrieffer, 
\emph{Phys.\;Rev.} {\bf 108}, 1175 (1957).

\bibitem{Cooper1956}
L.\,N. Cooper, 
\emph{Phys.\;Rev.} {\bf 104}, 1189 (1956).

\bibitem{Palanque-Delabrouille2015} N.~Palanque-Delabrouille \emph{et al.}, 
\emph{J.\;Cosmol.\;Astropart.\;Phys.} {\bf 2015}, 011 (2015).

\bibitem{Komatsu2011} E.~Komatsu \emph{et al.}, 
\emph{Astrophys.\;J., Suppl.\;Ser.} {\bf 192}, 18 (2011).

\bibitem{Navarro1995} J.\,Navarro, C.\,Frenk, and S.\,D.\,White, 
\emph{Mon.\;Notices\;Royal\;Astron.\;Soc.}, {\bf 275}, 720 (1995).
  
\bibitem{Geller1989} M.\,J. Geller and J.\,P. Huchra, 
\emph{Science} {\bf 246}, 897 (1989).

\end{thebibliography}
\end{document}